# IEEE Copyright Notice

# Identifying Oscillations Injected by Inverter-Based Solar Energy Sources

Chen Wang, Chetan Mishra, Kevin D. Jones,
R. Matthew Gardner

Dominion Energy
Richmond, VA, USA

Luigi Vanfretti

Rensselaer Polytechnic Institute
Troy, NY, US

*Abstract*— Inverter-based solar energy sources are becoming widely integrated into modern power systems. However, their impacts on the system in the frequency domain are rarely investigated at a higher frequency range than conventional electromechanical oscillations. This paper presents evidence of the emergence of an oscillation mode injected by inverter-based solar energy sources in Dominion Energy's service territory. This new mode was recognized from the analysis of real-world ambient synchrophasor and point-of-wave data. The analysis was performed by developing customized synchrophasor analytics tools deployed on the PredictiveGrid™ platform implemented at Dominion Energy. Herein, we describe and illustrate the preliminary analysis results acquired from spectrogram observations, power spectral density plots, and mode shape estimation. The emergence and propagation of this new mode in Dominion Energy's footprint is illustrated using a heatmap based on a proposed frequency component energy metric, which helps to assess this oscillation's spread and impact.

*Index Terms*-- Ambient data, Oscillation, Mode shape, Solar energy source, Synchrophasor.

I. INTRODUCTION

With the increasing integration of renewable energy sources (RES) into power systems, concerted efforts are been made to understand what impacts RES have on various aspects of system stability, including: voltage security [1], transient stability [2], [3], and small-signal stability [4], to name a few. Most of these studies are based on models with simulations and focusing on important problems that arise in conventional power system analysis. In terms of small signal stability analysis, [5] considers the impact of photovoltaics on system's electromechanical oscillations. However, the use of inverters (and their associated controls) for energy conversion in many RES can result in other oscillation modes (with higher frequencies than electromechanical modes) and, subsequently, influence the overall system behavior. To the knowledge of the authors, this emergent behavior has not been reported and studied in detail from utility field measurements.

In this paper, we report on a new oscillation mode found in ambient synchrophasor and point-on-wave data from various substations at Dominion Energy with solar energy sources (SES) penetration and/or close to SES points of common coupling. The key characteristic of this new mode is evidenced through spectrograms that closely track the mode's emergence during daylight hours. Dominion Energy has nearly 100% of its substations equipped with phasor measurement units (PMU, or synchrophasors) and digital fault recorders, providing a wealth of data that can be access through the cloud-based PredictiveGrid™ platform. Using this infrastructure, customized data analytics [6] have been implemented to perform the analysis herein. In this paper, we first demonstrate the discovery of this new mode using 48-hour spectrograms [7], and the identification of the actual mode frequency using higher reporting rate (RT) data and the Point-on-Wave (PoW) data. Based on the available data, we also conducted preliminary causality analysis, which informs on the likely source of the oscillation, the inverter controls. Next, we use power spectral density (PSD) plots to reveal the new mode's footprint. Lastly, results from mode-shape analysis and spatial visualization through a geographical heatmap provide further supporting evidence of the impact that the SES have on the system through the spread of this new mode. The main contribution of this work lies not only in the identification and report of this new mode injected by inverter-based solar sources, but also in the introduction of the data-driven methodology using existing frequency domain analysis tools for new oscillation identification based purely on PMU measurement data from the utility. The accuracy and effectiveness of the results and findings in this paper can be certified by these well-established tools, given that they have long been utilized in both diverse fields of engineering [8] and in power system monitoring [9].

II. SPECTRAL ANALYSIS

As a part of an ongoing survey for oscillations, spectrograms were obtained from voltage magnitude measurements throughout Dominion Energy's system for ambient operation conditions [10]. An unforeseen oscillation mode, detected around 8 Hz, was found in a wide group of locations at/or adjacent to substations with SES, as described next.

*A. Spectrogram at one representative Substation with SES*

Taking the example of a representative substation equipped with one inverter-based SES, the authors present the following 48-hour spectrogram, in Figure 1, calculated with the measurement data acquired from the PMU installed right upstream towards the PCC point of the SES in the substation. It is easy to notice the clear appearance of this mode with frequency around 8 Hz highlighted by the brighter color hue. The time points start from 07-01-2020 6:00 AM UTC-4. The spectrogram uses five minutes as the window of detrended data for Fast Fourier Transform (FFT). Within the two days investigated, this discovered mode started around 6:00 AM and ended around 8:00 PM daily, coinciding with sunrises and sunsets of summer days.

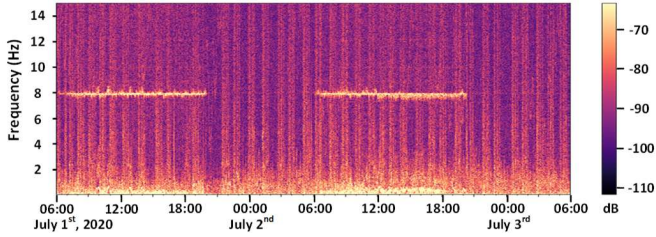

Figure 1. Summer day, *V* magnitude 48-hour ambient data spectrogram.

One interesting phenomenon is the activity of this mode in winter days as shown in Figure 2. Observe that even though the mode is present throughout the entire day, during daylight hours, the power of the corresponding frequency component is stronger and more concentrated around 8 Hz. One possible cause could be the different configurations of the SES controls in winter and summer days.

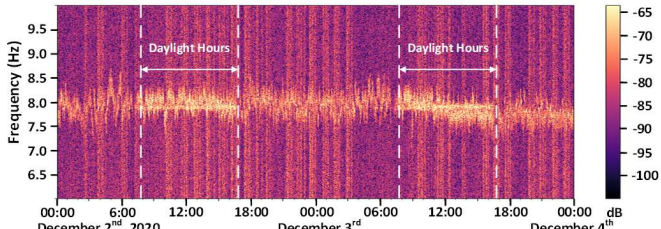

Figure 2. Winter day, *V* magnitude 48-hour ambient data spectrogram.

To further confirm the mode frequency, we also analyzed synchrophasor data with 60 samples per second (sps) of reporting time (RT). The 24-hour spectrogram of one typical voltage phasor starting from 12-19-2020 00:00 AM UTC-5 is shown in Figure 3. The result shows the existence of a stronger mode around 22 Hz, with the details labeled indicating similar patterns. Further analysis herein, reveals that the mode around 8 Hz is actually the result of aliasing from this actual oscillation mode around 22 Hz.

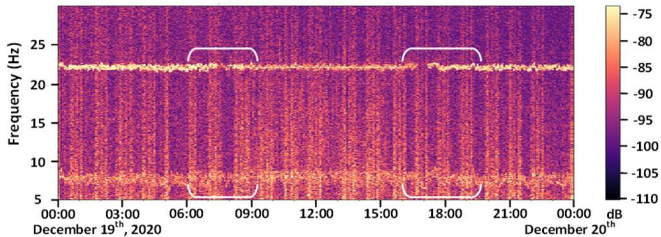

Figure 3. 24-hour spectrogram of winter day selected *V* magnitude with 60 sps.

To verify this, PSDs of the PoW data with 960 Hz sampling frequency were obtained, as shown in Figure 4. The methods used are the Short-time Fourier transform (STFT) [11] and demodulation using Hilbert's transform [12]. Note that

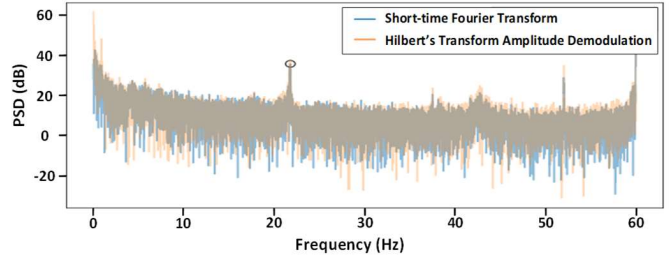

Figure 4. PSD of Point-on-Wave *V* magnitude data.

Dominion Energy has deployed a large fleet of digital fault recorders which has the capability of recording and storing (not streaming) with manually configured settings ambient PoW measurements for a predefined period of time. The accessibility to such high sampling rate data is thus limited and depend on substantial human intervention. In fact, to minimize communication costs, most utilities in North America use only 30 sps of RT for most PMU installations. Consequently, one of the major contributions of this work is to provide this analysis process as a guideline: *when analyzing unforeseen oscillation modes, verification using PMU data with higher RT and/or PoW data must be conducted. Once the frequency of the actual mode is identified, the use of PMU data at lower RTs has to be mapped onto the true mode frequency.*

### B. Causality Analysis

To further explain the discovered phenomenon, given clear correlation of the mode emergence in the spectrograms and daylight hours, we attempted to use the available data to analyze the causality between the observed mode and the inverter-based

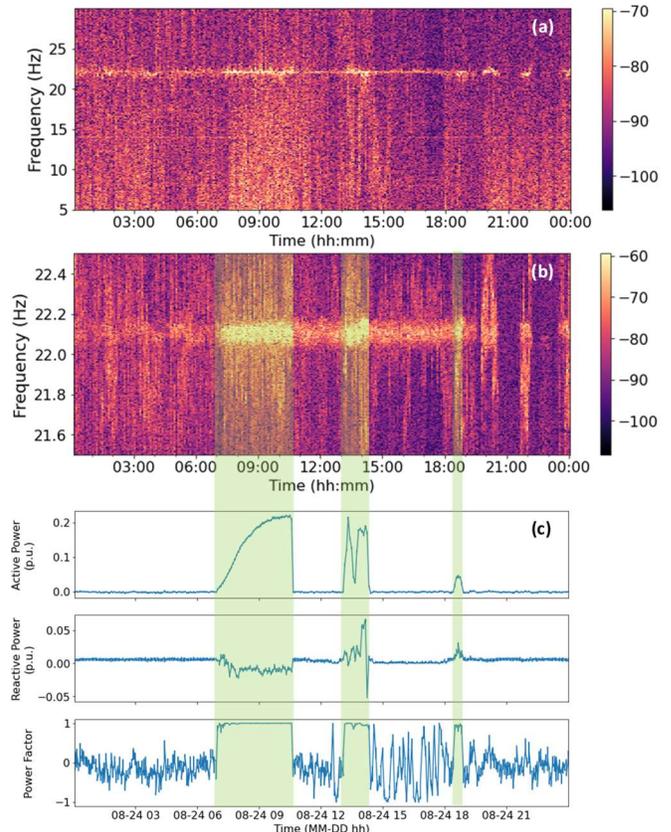

Figure 5. Spectrograms and power relevant plots at SES on Aug 24$^{th}$ 2021.

SES at the aforementioned substation. Another typical summer day (Aug 24th, 2021) was selected where the SES was switched on and off multiple times within 24 hrs., which could bring insights about the causality. In Figure 5, the spectrogram of 24-hour voltage magnitude measurements with 60 sps is shown in Figure 5 (a). The spectrogram shows that within this day, the 22Hz mode emerged in the morning and disappeared intermittently in the afternoon. Further, zooming into the designated frequency range from 21.5Hz to 22.5Hz in Figure 5 (b), the higher energy appearances of the mode can be easily observed (as highlighted by the green bands). It is also worth noting that besides the higher energy, there also exists a mode at the same frequency that is not product of the inverter's operation. This could be because of the other SESs in the system from the same vendor that were operating continuously, and the frequency component created by them propagated to the current observing point.

Furthermore, we investigated the relationship between the emergence of the 22Hz mode and PV control scheme, Figure 5 (c) shows the PV's time series active power output (P), reactive power output (Q), and the corresponding power factors (PF) of that specific day. It is to be noted that this calculation is based on the typical PMU data with 30 sps, given that we are attempting to map the whole 24 hours steady state operation conditions with frequency domain results and there is no need for more detailed granularity. Looking at the green bands that show when the mode's energy is high, it can be easily seen that the mode energy increased along with the PV's power output, moreover, it can be observed that the PV is operating under constant P and constant power factor (PF) control scheme. This clearly shows the relevance between the mode emergence and the P-PF control scheme.

To further demonstrate, we provided following scatter plots of the operation points on the same day in Figure 6 and Figure 7 indicating that the mode energy was higher when the PV injects active power into the bulk system (the green dashed lines are used for visual inspection to better differentiate clusters). Figure 8 directly shows that when the PV plant operated in constant PF control scheme the energy of the designated mode was clearly stronger. Similar mechanism was explained in [13]. All this information coincides with authors' hypothesis, as stated in the previous paragraph.

Despite the clear relationships demonstrated by these distributions of operation points, to confirm the causality the authors believe real-world field tests are necessary. The authors

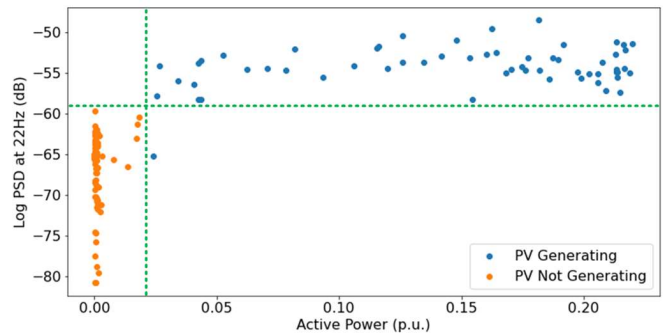
Figure 6. Relationship between active power and instantaneous voltage magnitude measurements 22Hz mode energy in 24 hours.

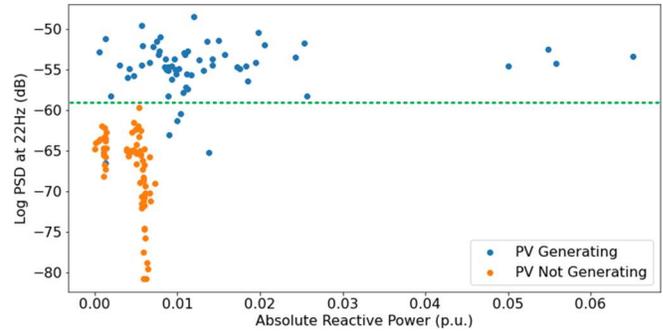
Figure 7. Relationship between absolute reactive power output and instantaneous voltage magnitude measurements 22Hz mode energy in 24 hours.

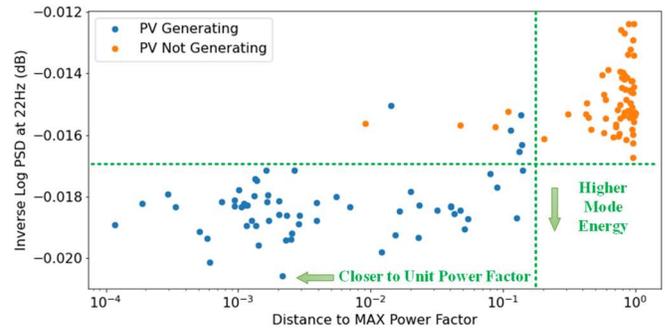
Figure 8. Relationship between power factor and instantaneous voltage magnitude measurements 22Hz mode energy in 24 hours.

are proactively seeking the opportunity of such tests and detailed models of relevant devices. A more rigorous root cause analysis will be performed and reported in future works after field tests have been conducted.

III. SYSTEM-WIDE IMPACT ANALYSIS

Spectral analysis was also conducted on measurements from substations distributed over Dominion's service territory to reveal the system-wide impacts of the observed mode. It is to be noted that, given limited access for higher RT (60 sps) PMU data and PoW data, the reminder of this paper uses the 8 Hz oscillation estimates as a proxy to assess the spread and impact of the actual 22 Hz mode.

A. PSDs for System-wide Voltage Phasor Measurements

Another phenomenon observed is this oscillation mode appears more significantly in voltage magnitude measurements, so we conducted a survey using PSD plots of voltage magnitudes on all PMUs throughout Dominion's system to find the propagation of this mode within the system's topology. Figure 9 shows the signals' PSD plots containing this frequency

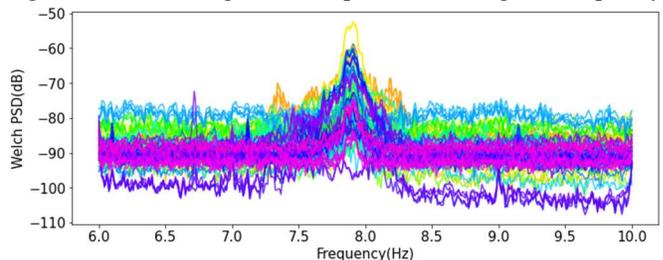
Figure 9. PSD plots of system-wide $V$ magnitude measurements with 20-min detrended data using the Welch's method.

component using Welch's method. The measurements selected here are ambient synchrophasor data spanning a 20-minute window starting at 07-01-2020 11:50 UTC-4. Each one-minute segment is used for the FFT with no overlapping. The 97 voltage measurements included in this plot are distributed among 25 substations. Clearly, most of the signals show strong power density around 8 Hz, however, it better defined certain locations, this is an indication of how the mode decays as it propagates [14]. Using Yule-Walker's algorithm the estimated mode frequency was found at 7.89 Hz.

Further analysis of the data reveals two scenarios for the emergence of the mode. Figure 10 presents the most common scenario (Scenario 1), where this new mode only appears in voltage phasors (magnitudes-VPHM and phase angles-VPHA), but not in current phasors (IPHM and IPHA). Calculated active and reactive power data are also not exposed to the mode here.

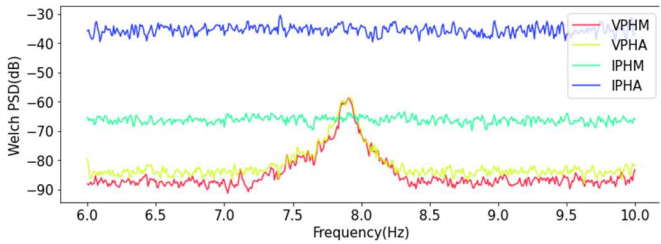
Figure 10. Scenario 1: PSDs of one *V* and *I* phasor pair with 20-min detrended data using the Welch's method.

Under Scenario 2, both voltage and current phasors reflect this mode, see Figure 11. Active and reactive power PSD plots also contain this frequency component. Although Scenario 2 is rare, it was mostly observed at substations with SES penetration.

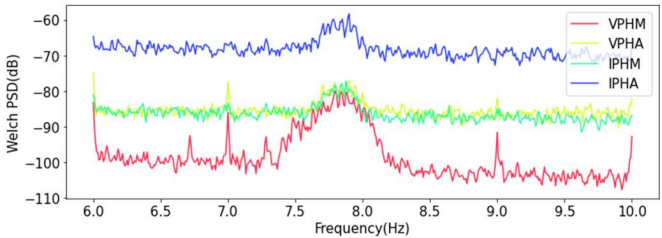
Figure 11. Scenario 2: PSDs of one *V* and *I* phasor pair with 20-min detrended data using the Welch's method.

We hypothesize that the difference in the two scenarios discovered is due to different control strategies used by various SES, as discussed in the previous section.

### B. Mode Multiplicity and Mode Shape

Because this discovered mode appears in almost 100 different PMU streams, it was necessary to determine if there was more than one mode within the observed frequency range. We conducted a mode multiplicity analysis using the method introduced in [15], which applies singular value decomposition on the cross spectral density matrices around the frequency of interest. By plotting the most significant singular values, one can differentiate various modes in the curve's peaks and further find corresponding mode shapes with singular vectors. Figure 12 illustrates the top three significant singular values between 7 to 9 Hz. Based on these results, it was possible to confirm that

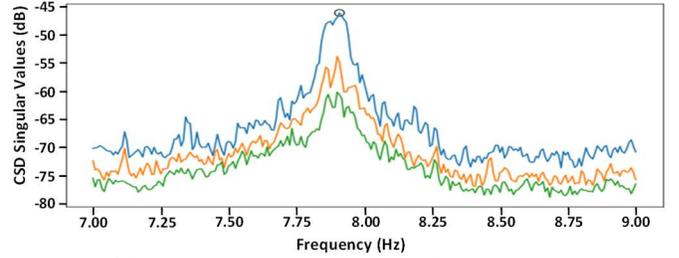
Figure 12. Top three singular value plots to find mode multiplicity.

there was only one mode because the peaks and shapes of the three curves coincide with each other, i.e. no mode multiplicity.

Given this result and the fact that the discovered mode oscillates among multiple substations, determining the mode shape was of interest for analysis because it could provide insights on the mode source(s). The mode shape was obtained with a Yule-Walker mode meter at the frequency of 7.89 Hz. The estimated mode shape plot is shown in Figure 13. The color code for each stream is the same as in the PSD plot in Figure 9. The colors are assigned based on the geographical locations of the measurements in the network's topology. The proximity in color hue in the maps indicates that the corresponding locations are at adjacent substations and/or substations connected by shorter transmission lines. As shown in the mode shape, this mode appears most prominently in the Central Region. Substations in the North and East Regions are more likely to swing along with the Central Region, while substations in the South and West Regions swing against it. Exploring the details of the mode's geographical distribution is necessary to locate the potential source and clearly demark the spatial spread.

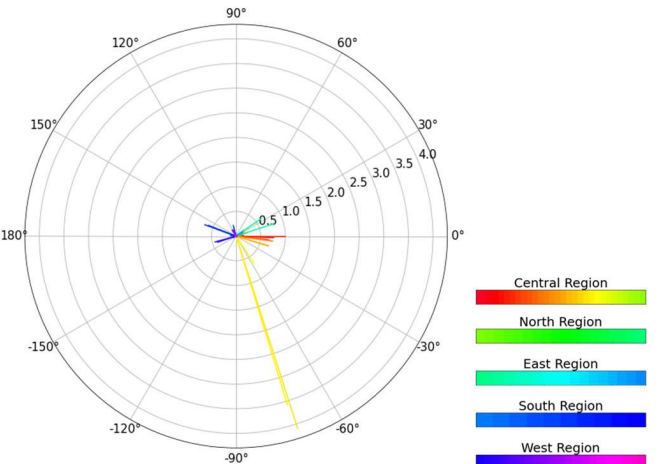
Figure 13. Mode shape at 7.89 Hz of selected voltage magnitude.

### C. Mode Geographical Spread

To better quantify the geographical spread of this mode, we propose to use the percentage of the mode area under the PSD curve over the area within certain frequency range as a metric of the mode energy in each measurement. The calculation is shown in (1).

$$E_{percent} = \frac{\int_{f1}^{f2}[psd(f) - \text{trend}(psd(f_{nm}))]df}{\int_{f1}^{f2}[psd(f) - psd_{min}]df} * 100 \quad (1)$$

where $psd(f)$ indicates the PSD curve value at frequency $f$, $[f_1 - f_2]$ is the range of frequencies of interest, $trend(psd(f_{nm}))$ is the trend of the PSD outside the mode part, $psd_{min}$ is the minimul power density within concerned frequency range. Here we select $[5, 11]$Hz to be the range of interest to avoid electromechanical components. Figure 14 exhibits the resultant heatmap based on the proposed metric (geographical contextualization is omitted due to confidentiality). The Central Region is labeled on the map. The "hotter" color hue in this region confirms the mode's extraordinary energy there. The outstanding density of SESs installation (both utility owned and distributed) in this region is able to explain this phenomenon. The other highlighted areas left and right of the Central Region correspond to the West and East Regions. For readers' reference, the geographical distance is around 70 miles between the centers of the high energy areas in Center Region and West Region respectively. All this information confirms the mode shape results.

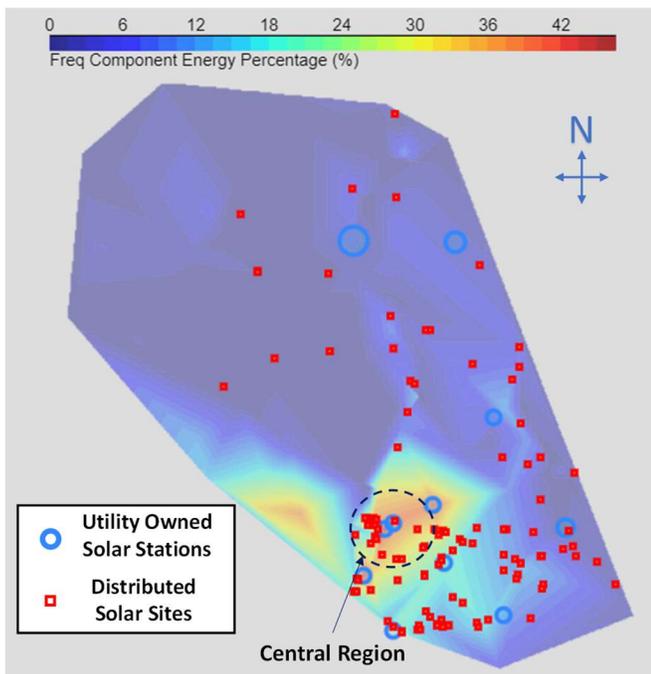

Figure 14. Discovered mode distribution throughout Dominion footprint.

## IV. Discussion and Conclusion

In this paper, an oscillation mode injected by SES's in the Dominion Energy service territory has been reported. The daytime appearance of this discovered mode supports the strong connection between this oscillation and the SES in the system. Preliminary causality analysis results also coincide with the hypothesis the authors proposed where the emergence of this mode is highly likely caused by the P-PF control scheme of the SES. The geographical plots help further strengthen this connection. Thanks to the synchrophasor data available in the cloud-based platform, PredictiveGrid<sup>TM</sup>, customized synchrophasor analytics were implemented to conduct a system-wide spectral analysis. The results show that the installation of SES, whether they are utility-owned or distributed sources, have a strong correlation with the mode energy at the southern part of the network. This matches the dense spatial distribution of inverter-based SES integrated in that area. Future work will perform model-based analysis and field tests to solidly determine the control mechanisms in the inverter-based SES that give rise to this new mode.